\documentclass[aps,prl,showpacs,onecolumn,superscriptaddress,floatfix]{revtex4-2}
\usepackage[utf8]{inputenc}

\usepackage{verbatim}
\usepackage{amsmath}
\usepackage{mathtools}
\usepackage[version=3]{mhchem}

\DeclarePairedDelimiterX\braket[2]{\langle}{\rangle}{#1 \delimsize\vert #2}

\usepackage[pdftex]{graphicx}
\usepackage[separate-uncertainty=true]{siunitx}
\DeclareSIUnit{\rpm}{rpm}

\usepackage[colorlinks=true,urlcolor=blue,linkcolor=blue,citecolor=blue]{hyperref}
\usepackage[usenames,dvipsnames]{xcolor}
\usepackage[T1]{fontenc}
\usepackage{placeins}
\usepackage{nicefrac}
\usepackage{braket}
\usepackage{pdfcomment}
\usepackage{xcolor}
\usepackage{braket}
\usepackage[normalem]{ulem}

\usepackage{gensymb}

\usepackage{lineno}

\usepackage{titlesec}
\titleformat{\section}{\normalfont\normalsize\bfseries\centering}{\thesection}{1em}{}
\titleformat{\subsection}{\normalfont\normalsize\bfseries\centering}{\thesubsection}{1em}{}
\titleformat{\subsubsection}{\normalfont\normalsize\bfseries\centering}{\thesubsubsection}{1em}{}

\begin{document}

\title{Spin-quenching in molecule-transition-metal-dichalcogenide heterostructure through inverse proximity effect}
%
%

%
\author{Swagata~Acharya}\email{Swagata.Acharya@nrel.gov}
\affiliation{Materials, Chemical, and Computational Science Center, National Renewable Energy Laboratory, 15013 Denver West Pkwy, Golden, CO 80401}
\author{Dimitar Pashov}
\affiliation{King’s College London, Theory and Simulation of Condensed Matter, The Strand, WC2R 2LS London, UK}
\author{Daphn\'e Lubert-Perquel}
\affiliation{Chemistry and Nanoscience Center, National Renewable Energy Laboratory, 15013 Denver West Pkwy, Golden, CO 80401}
\author{Mark~van Schilfgaarde}
\affiliation{Materials, Chemical, and Computational Science Center, National Renewable Energy Laboratory, 15013 Denver West Pkwy, Golden, CO 80401}

\author{Justin C. Johnson}
\affiliation{Chemistry and Nanoscience Center, National Renewable Energy Laboratory, 15013 Denver West Pkwy, Golden, CO 80401}%

\begin{abstract}
A functional heterostructure lies at the heart of integrated circuitry involved in quantum photonics, optoelectronics, neuromorphic computing, spintronics,
and straintronics. Over the past few years several heterostructures involving two-dimensional (2D) magnets and (nonmagnetic) transition metal dichalcogenides (TMDs) have been explored. The electron and hole wavefunctions are nearly atom-local in the 2D magnets while in the nonmagnetic TMDs they are delocalized over several unit cells. In a heterostructure made out of these two, a proximity-induced magnetic inter-layer exciton can emerge, whose energy differs by $\sim$20-30 meV of the excitons confined to one or the other layer, and is also two orders of magnitude darker, making it both hard to detect and functionalize. In this work, using a high fidelity \textit{ab-initio} many-body diagrammatic approach we show that the degree of functionality can be significantly enhanced in a transition-metal molecule-TMD interface. We show that the molecular exciton has charge transfer character and is extended, in strong contrast to the localized Frenkel excitation in a 2D magnet.  Moreover, we show the degree of localization and molecular magnetic moment can be tuned rather easily, through varying the molecular orientation relative to the TMD.  Variations in orientation modifies the proximity of the TMD to the magnetic ion, thus changing the screening and providing a pathway through which the proximity effect can quench the spin moment of the magnetic ion.  This effective inverse proximity effect is able to tune the energies, spin states and brightness of the molecular and the inter-layer magnetic excitons, a mechanism that is absent in 2D-magnet/TMD heterostructures. We also pin down the conditions under which the inter-layer exciton can be well split off and made brighter compared to its neighboring intra-layer exciton making it highly attractive for devising protocols reliant on probing and manipulating magnetic excitonic states. 
\end{abstract}

\maketitle
\section{Introduction}

Heterostructures of transition metal dichalcogenides (TMDs) have emerged as a primary vehicle for integrated circuitry in
quantum photonics, optoelectronics, neuromorphic computing, spintronics, and straintronics
~\cite{pham20222d,kroemer2005heterostructure,Zutic2004:RMP,liu2016van,geim2013van,chhowalla2013chemistry,huang2017layer,seyler2018ligand}.  New proximity-induced quantum phenomena
emerge~\cite{zuticpe1,montero2002nanostructures,deutscher2018proximity}, which enhance the ability to tune the heterostructural properties in situ.  These proximity effects differ from traditional applications~\cite{Hauser1969:PR}, as the aim is to use two-particle properties (excitons) as the primary vehicle for the circuits.  Heterostructures of TMDs and hexagonal boron nitride (hBN)~\cite{yoon2022charge,ding2018understanding,epstein2020highly,uchiyama2019momentum,mueller2018exciton,nerl2024mapping,yagodkin2022extrinsic,li2020resonant,qian2022nonlocal,hotta2020enhanced,PhysRevLett.123.067401,zhao2022exciton,21gh-fxyh,rosati2025impact} are a very popular route to tunable excitons via a proximity effect.  A series of excitons
can be realized over the visible and ultraviolet window, several of which have large binding energies ranging up to a
few 100's of meV.  These excitons are nonmagnetic and so are the excitons observed in the TMD heterobilayers~\cite{yu2015anomalous,wu2018theory,baranowski2017probing,rivera2018interlayer} which implies that often it is the pseudo-spin and/or valley degrees of freedom that are tuned in such heterostructures in absence of an explicit spin degree of freedom. More recently, TMDs are functionalized with heterostructures of magnetic van der Waals (vdW) films and splitting of the non-magnetic TMD excitons in the proximity of a magnetic field genereated by the vdW magnets is used to access the microwave and
infrared resonant and non-resonant energy windows~\cite{norden2019giant,ge2022enhanced,seyler2018valley,zhao2024magnetic,heissenbuttel2021valley,hu2020manipulation,zollner2019proximity,garcia2025conductance,ali2025magnetic}.
  These vdW materials host a spectrum of excitonic transitions~\cite{pollini1970intrinsic,grant1968optical,dillon1966magneto,zhang2019direct,ye2014probing,qiu2013optical,acharya2022real,datta2025magnon,shao2025magnetically,grzeszczyk2023strongly} which
are often well split off from the band edges and are not contaminated by the continuum of one-particle excitations.  In
either case, the two functionalities --- integrated circuitry at the nanoscale, and the ability to tune excitons in situ
--- promise a new era of efficient nanoscale quantum integrated
circuitry~\cite{fiori2014electronics,jariwala2014emerging,zhang2015ultrathin}, superseding traditional semiconductor
applications in logic, memory, sensing, and coherent emission with nanoscale devices.

That being said, in-situ control of these properties through doping or alloying remains a daunting task.  Another drawback of
these approaches is that the range of tunability for the excitonic states is modest - the magnetic splittings of the (otherwise non-magnetic TMD) excitons are often less than 1 meV, the emergent interlayer excitons are significantly dark and they often reside very close to the intra-layer excitons which are brighter making detection and control of the emergent states difficult.  Here we consider an
alternative that shows promise for a greater degree of tunability, namely a heterostructure of a TMD and a molecule with
a (magnetic) transition metal ion.  We show that there is an inverse proximity effect (the TMD modifying the molecule)
which has a marked effect on the molecule's magnetic state.  In the 2D magnet/TMD heterostructure the proximity effect is
unidirectional where the large frozen spin moments of the magnet generate a proximity field and split the otherwise nonmagnetic
excitonic states of the TMD. The functionality of such heterostructures, primarily relies on the ability to induce magnetic field
and split the TMD excitons by desired amounts. 

\begin{figure}[ht]
\includegraphics[width=16cm]{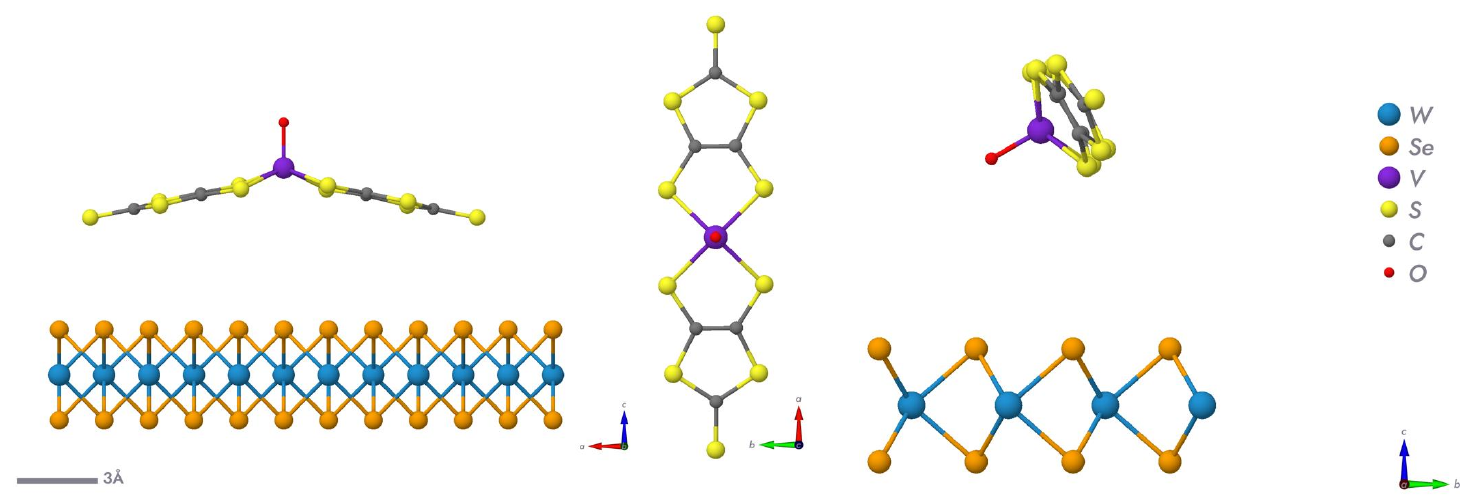}
    			\caption{The simulated crystal structures of the different VO(dmit)$_{2}$-TMD heterostructures are shown. The molecule (shown in the middle separately) is rotated at different angles with respect to the TMD layer. Two cases are shown with the angle between the layers respectively at  0\degree and 120\degree. RGB is used to represent the standard $abc$ crystallographic directions. 
                }
	\label{fig:struc}
\end{figure}

In the present work we show that the situation changes rather remarkably in a (magnetic) molecule/TMD heterostructure.
Screening induced in the molecule by the TMD is strong enough to produce an inverse proximity effect by which the
relative level alignments between the TMD and the molecule, the \textit{d-d} splitting of the molecular states and most
importantly the magnetic moment of the molecule can be tuned at will. To this end, we explore various heterostructure
configurations with varying interfacial angles (and magnetic TM sites) between the molecule and TMD. We predict
interfacial geometries that can be synthesized in a controlled manner to realize this unique inverse proximity effect
and access a full spectrum of spin states in the molecule starting from the  high spin state to
a fully spin quenched state. The organic molecules with partially filled TM ions realize charge transfer
excitons with spatially extended wavefunctions and, hence, are much softer compared to the spin-frozen TM sites from the vdW
magnets. As a result, molecule-TMD heterostructures provide a unique knob to tune the magnetic moments, valence and
conduction states and their energies, and the brightness of the interfacial excitons, a mechanism lacking in the 2D vdW magnet/TMD interfaces.  

Various combinations of TMD/molecule heterostructures with potential for strong interlayer coupling have recently been studied, both experimentally and computationally~\cite{ulman_organic-2d_2021, dziobek-garrett_excitons_2024}, including some that exhibit orientation-dependent exciton hybridization~\cite{bartlam_orientation_2024, fu_resonantly_2025}.  However, the specific interplay between the TMD exciton and the proximal spin hosted on the molecule has not been explored despite the potential for exquisite control of the resulting magnetic behavior.  As one realization of this phenomenon, we investigate here the molecule [VO(dmit)$_{2}$], with dmit = 1,3-dithiole-2-thione-4,5-dithiolate, as a model due to the vanadyl spin exhibiting long quantum coherence times~\cite{atzori_quantum_2016}. The $T_{2}$ relaxation of > 1 $\mu$s  at room temperature promises future application as a qubit for quantum sensing or computation, including extended lifetime to milliseconds under certain conditions and optical readout~\cite{zadrozny_millisecond_2015, fataftah_trigonal_2020}. The compound can be conveniently diluted in an identical structure but with a diamagnetic TM, such as Mo that we employ here.  The [VO(dmit)$_{2}$] compound is related to vanadyl phthalocyanine (VOPc), a commonly studied molecular semiconductor that we recently reported has modulated interfacial coupling with WSe$_2$ through molecular orientation control~\cite{lubert-perquel_modulating_2024}. The molecular orientation was locked in during the sample preparation process, through thickness control where the initial layer of VOPc was face-on to WSe$_2$, while crystallization forces in thicker layers pulled the molecules to adopt an edge-on orientation. [VO(dmit)$_{2}$] is expected to possess similar properties to VOPc but with fewer atoms in the scaffold surrounding the vanadium (IV) center, making it computationally feasible to investigate within our higher level diagrammatic many-body theory.

\begin{figure}[ht]
\begin{center}
    \includegraphics[width=14cm]{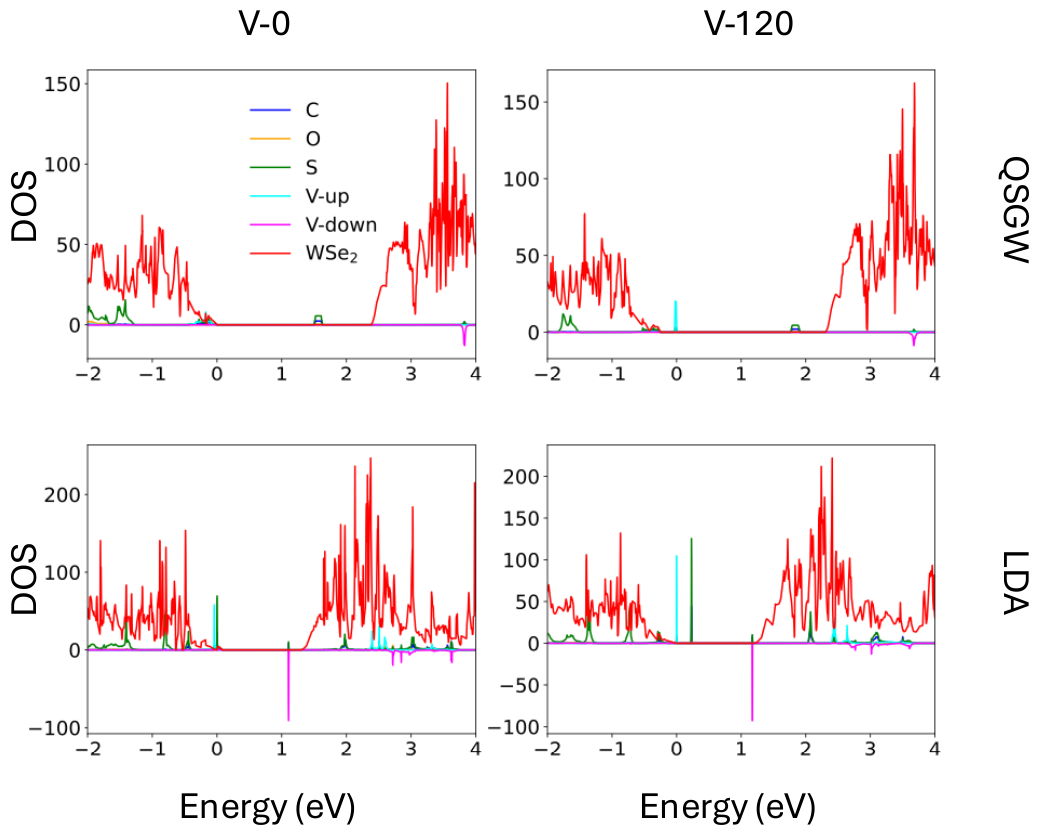}
\includegraphics[width=14cm]{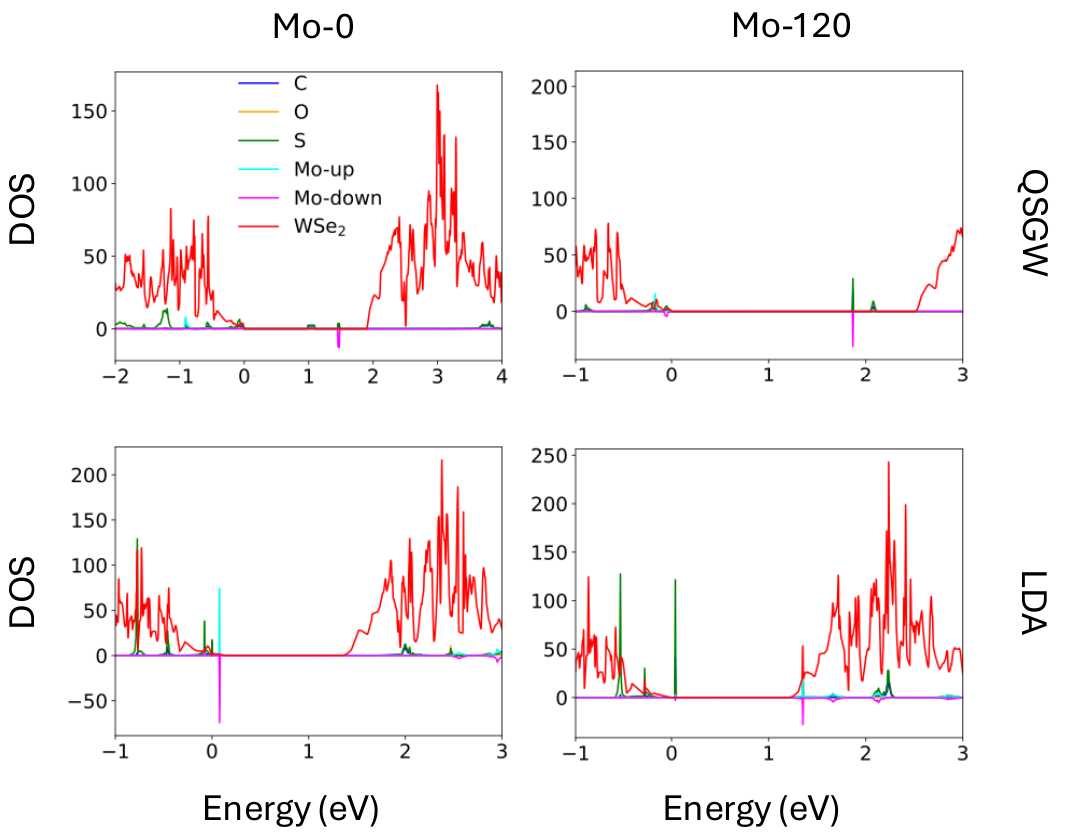}  
\end{center}\caption{Partial density of states (DOS) of the different heterostructures are shown. The DOS are shown from two levels of the theory:  LDA and $\mathrm{QS}G\hat{W}$. Partial DOS for four different heterostructure configurations are projected on the molecular layer containing Carbon, Oxygen, Sulfur, Vanadium/Molybdenum (transition metal) and the WSe$_{2}$ layer and the $\uparrow$ and $\downarrow$ spins of the transition metal element. The change in bandgap, and especially $\mathrm{QS}G\hat{W}$'s reordering/realignment of the frontier orbitals has a pronounced effect on the $dd$ splittings and the excitons.}
	\label{fig:pdos}
\end{figure}

\section{Results}
We discuss four crystal structures in the following; two molecules with V and Mo as TM sites and for each molecule two angles (0\degree\ and 120\degree) between VO(dmit)$_{2}$ and WSe$_{2}$ layers are simulated. The distinct configurations are illustrated in Fig.~\ref{fig:struc}. For convenience we refer to these four different structures as V-0, V-120, Mo-0 and Mo-120 in the rest of the paper. Electronic and excitonic properties are computed using local density approximation (LDA), many-body perturbation theory in the quasi-particle self-consistent $GW$, QS$GW$~\cite{qsgw} and $\mathrm{QS}G\hat{W}$~\cite{Cunningham2023} from the open-source electronic structure suite  Questaal~\cite{questaal_paper}. The static quasi-particle self-energy $\Sigma^0(\mathbf{k})$ is generated on a 3$\times$6$\times$1 k-mesh and the dynamical self-energy $\Sigma(\mathbf{k},\omega)$ is generated on a 2$\times$2$\times$1 k-mesh. In each iteration of the QS$GW$ ($\mathrm{QS}G\hat{W}$) self-consistency cycle, $\Sigma^0(\mathbf{k})$ and the charge density are both updated.  Iterations continue until self-consistency in both is reached. The convergence threshold for the self-energy is set at $2{\cdot}10^{-5}$ Ry for the self-energy. The unit cell contains 90 atoms, 18 atoms from the molecule and 72 atoms from the WSe$_{2}$ layer.  Note that the $\mathrm{QS}G\hat{W}$ contains the excitonic vertex missing from the QS$GW$ level of the theory. The excitons are computed within $\mathrm{QS}G\hat{W}$ by including ladder diagrams via Bethe-Salpeter-equations (BSE). The BSE optical vertex is computed using 5 valence and 4 conduction states while the self-energy and charge density are made out of all electronic states. 
\begin{table}[h]
    \centering
    \begin{tabular}{|c|c|c|c|c|}
    \hline
    System  &  LDA E$_{g}$ (eV) & LDA Magnetic moment ($\mu_B$) (local, bonds) & $\mathrm{QS}G\hat{W}$ E$_{g}$ (eV) & $\mathrm{QS}G\hat{W}$ Magnetic moment ($\mu_B$) (local, bonds) \\
    \hline
    V-0  & 0.048 & 1.10 (0.72, 0.38) &  1.60  & 1.09 (0.75, 0.34) \\
    V-120  & 0.230 & 1.10 (0.72, 0.38) & 1.76 & 0.92 (0.60, 0.32) \\
    Mo-0  &  0.02 &  0.04 (0.02, 0.02) & 1.0 & 0.89 (0.38, 0.51) \\
    Mo-120 & 0.036  & 0.02 (0.01, 0.01) & 1.6  & 0.47 (0.2, 0.28) \\   
    \hline
    \end{tabular}
    \caption{Band gaps E$_{g}$, the magnetic moments and their distribution over the atomic and the bonded components from LDA and $\mathrm{QS}G\hat{W}$ for four different heterostructure variants are reported.}
    \label{tab:bandgap}
\end{table}

In pristine WSe$_{2}$ (and also the V-0 heterostructure, as seen from the partial DOS in Fig.~\ref{fig:pdos}) the bandgap comes out at 2.3 eV from $\mathrm{QS}G\hat{W}$ which is same as the experimentally reported value~\cite{he2014tightly,hanbicki2015measurement}. At the LDA level, the WSe$_{2}$ band gap is $\sim$1.5 eV, the underestimate being a well known limitation of LDA. Perhaps even more important in this context, LDA has a severe shortcoming in predicting both the alignment and the magnetic moment originating from the V-3$d$ occupation. This is the case since density-functionals make, by construction, the same potential for all electrons, and do not distinguish between $s$, $p$ and $d$ states. When correlations are strong in one channel, for example in partially filled, localized 3$d$ states, errors in LDA are larger compared to delocalized \textit{sp} molecular orbitals. This has an important bearing on the relative alignment between the S-$p$, V-3$d$ and WSe$_{2}$ states. In LDA, the V-3$d$ spin-split occupied (hole like $d^\uparrow$) and unoccupied (electron like $d^\downarrow$) states sit in the middle of the gap produced by the WSe$_{2}$ states. The valence frontier orbital has S-$p$ character and sits above the WSe$_2$ continuum , while the V-$d^\uparrow$ state appears $\sim$50 meV below with the V $d^\uparrow$-$d^\downarrow$ split by $\sim$1.2 eV.  $\mathrm{QS}G\hat{W}$ is very different. the S-$p$ orbital is pushed well below the WSe$_2$ continuum and the $d^\uparrow$-$d^\downarrow$ splitting is $\sim$4 eV, placing the $d^\downarrow$ state nearly 1.5 eV above the unoccupied WSe$_{2}$ continuum. That the $d^\uparrow$-$d^\downarrow$ splitting is often too small in LDA is well known, and occurs because correlations are underestimated. This splitting is an intuitive way to extract the effective Hubbard correlations parameters from theory. For example, in V-0 the magnetic moment of V is 1.10 $\mu_{B}$ in LDA and 1.09 $\mu_{B}$ in $\mathrm{QS}G\hat{W}$ (see Table~\ref{tab:bandgap}). So although the spin moments are essentially similar in two very different levels of the theory the $d^\uparrow$-$d^\downarrow$ splitting is very different. In a first order approximation, the $d$ band centers are split by $\pm$ $I{\cdot}M$, where $I$ is the Stoner parameter and $M$ is the magnetic moment. This implies an effective $I$ roughly 3 times larger in $\mathrm{QS}G\hat{W}$ compared to LDA. However, such analysis connecting the moment to the $d^\uparrow$-$d^\downarrow$ splitting should be valid for the atomic component of the moment and not for the total moment (atomic+ligand components). Delving deeper in to the internal structure of the magnetic moments, we find that of the 1.1 $\mu_{B}$ total moment, only about 2/3 resides inside of the atomic sphere ($\sim$0.72 $\mu_{B}$), while the rest of the moment is on the ligands, suggesting a fairly diffused character of the moment.
\textcolor{black}{(This is in keeping with V being an early transition metal, with a shallow \textit{d} state.)}
Using LDA values for $M$ and $d^\uparrow$-$d^\downarrow$ splitting, 0.72 $\mu_{B}$, $I$ is estimated to be $\sim$0.83 eV, which is similar to the Stoner parameter (1\,eV) typically used for 3$d$ transition metals. \textcolor{black}{At the ${\mathrm{QS}G\hat{W}}$ level the moment is sensitive to structure (Table~\ref{tab:dd}), a consequence in part of modifications to screening in the molecule.  This softness in the moment can further tuned by the ligand-metal charge transfer characters.} Also, 3$d$ states are more localized than its 4$d$ counterparts and this suggests that replacing V with Mo may provide better knobs to quench the spin moment in the heterostructures.

\begin{table}[h]
	\centering
	\begin{tabular}{|c|c|c|c|c|c|c|}
		\hline
		System  &  LDA $\Delta_{dd}$ (eV) & local LDA moment ($\mu_B$) & I$_{LDA}$ (eV) & $\mathrm{QS}G\hat{W}$ $\Delta_{dd}$ (eV & local $\mathrm{QS}G\hat{W}$ moment ($\mu_B$) & I$_{\mathrm{QS}G\hat{W}}$  \\
		\hline
		V-0  & 1.2 & 0.72 & 0.83 &  4 & 0.75 & 2.67  \\
		V-120  & 1.2 & 0.72 & 0.83 & 3.6 & 0.6 & 3.0 \\
		Mo-0  & $\sim$0 &  0.02 & $(-)$  & 2.4 & 0.38 & 3.15 \\
		Mo-120 & $\sim$0  & 0.01  & $(-)$ & 1.8 & 0.20 & 4.5 \\   
		\hline
	\end{tabular}
	\caption{The effective Stoner parameter $I$ (extracted from the theory), local atomic component of the magnetic moment and the $d^\uparrow$-$d^\downarrow$ splittings $\Delta_{dd}$ from LDA and $\mathrm{QS}G\hat{W}$ for the different heterostructures.}
	\label{tab:dd}
\end{table}

For V-0, the atomic components of $M$ are calculated to be 0.72 and 0.75 $\mu_{B}$ in LDA and $\mathrm{QS}G\hat{W}$ respectively. At the LDA level, 0.72 $\mu_{B}$ survives in V-120, and the $d^\uparrow$-$d^\downarrow$ splitting also remains unchanged from V-0. Notable differences are observed at the $\mathrm{QS}G\hat{W}$ level, though. The atomic component of the magnetic moment gets significantly screened in V-120 and reduces by 20\% to 0.6 $\mu_{B}$, proportionally reducing the $d^\uparrow$-$d^\downarrow$ splitting consistent with the $\pm$ $I{\cdot}M$ rule. This suggests that as the orientation of the magnetic element, V ion in this case, changes with respect to the WSe$_{2}$ layer important changes happen to the local (and non-local) part of the correlations that is captured by the self-energy in $\mathrm{QS}G\hat{W}$, an effect that the LDA fails to do. More importantly this provides an intuitive way to traverse through the spin-space and access different spin moments in these heterostructure configurations. Note that, this also provides a pathway to tune and control various Type-I to Type-II heterostructure configurations by changing the spin moments. For example, if the atomic part of the V spin can be quenched enough, say smaller than $\sim$0.4 $\mu_{B}$ (total moment less than $\sim$0.7 $\mu_{B}$)  the $d^\uparrow$-$d^\downarrow$ splitting can become smaller than the WSe$_{2}$ band gap of 2.3 eV changing the alignment from Type-I to Type-II.

\begin{figure}[ht]
 \includegraphics[width=16cm]{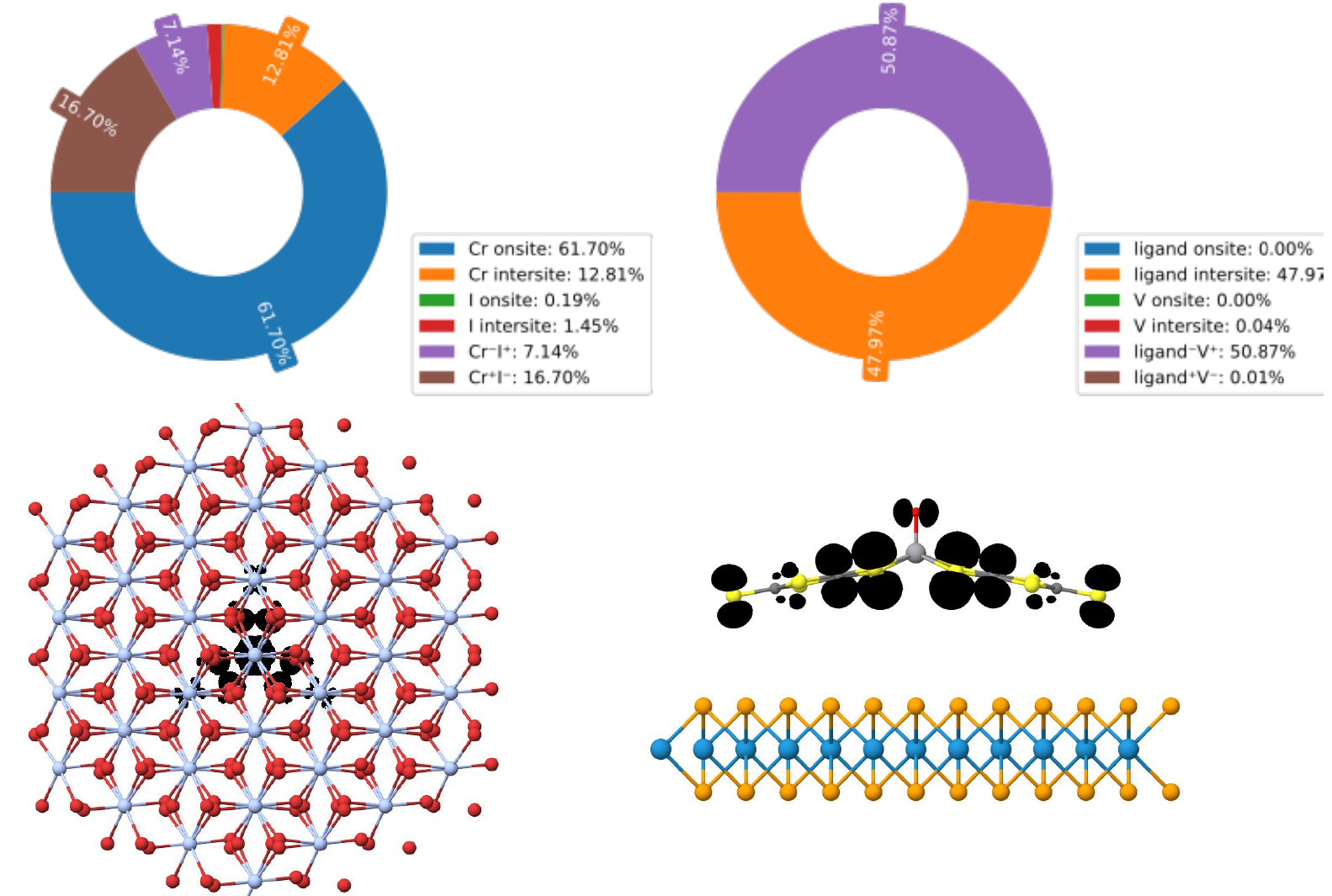}
\caption{\textcolor{black}{Contrast between a Frenkel exciton (CrI$_3$, left) and a charge-transfer exciton (VO(dmit)$_{2}$, right). The bottom parts depict the spatial (real-space) extent of the excitons. 
For CrI$_3$ most of the exciton resides on a single Cr site (electron and hole on the same site). This is shown as blue in the donut plot in the top left panel.  This magnetic exciton is highly localized reflecting its Frenkel character, extending only about $\sim$0.5 nm in real space.  The charge-transfer exciton on the right extends over the entire molecule $\sim$2.5 nm and has a vanishingly small onsite character. This exciton has two roughly equal components: V hole coupled to ligand electrons (purple) and ligand hole coupled to ligand electrons (orange).}}
	\label{fig:frenkelvsct}
\end{figure}

With Mo-0 and Mo-120, the situation is even more intriguing. At the LDA level the extended 4$d$ states of Mo are unable to pick up any spin moment and hence there are no $d^\uparrow$-$d^\downarrow$ splittings (see Fig.~\ref{fig:pdos}). However, this is not a physical spin-quenching mechanism, rather has its roots in the limitation of the LDA theory in describing the magnetic correlations of the TM sites. $\mathrm{QS}G\hat{W}$ is able to find magnetic moments of 0.89 $\mu_{B}$ and 0.47 $\mu_{B}$ respectively in Mo-0 and Mo-120. Only $\sim$40\% of the observed moments (0.38 and 0.20)  come from the local atomic component while the rest $\sim$60\% is diffused over the bonds, consistent with the extended character of the 4$d$ orbitals. The $d^\uparrow$-$d^\downarrow$ splittings in Mo-0 and Mo-120 are respectively 2.4 eV and 1.8 eV, which are consistent with the $\pm$ $I{\cdot}M$ rule (moderate deviations are due to the interplay of crystal-field effects and the extended character of 4$d$ orbitals). Further, the WSe$_{2}$ \textcolor{black}{virtual state continuum} can be observed at different energies, at 2 eV and 2.5 eV respectively in Mo-0 and Mo-120, suggesting stronger hybridization between the Mo-4$d$ and WSe$_{2}$ states, a mechanism mostly absent in V-3$d$ case. Together, they suggest that while Mo has similar spin quenching mechanism as that of V, the degree of tunability is significantly enhanced in Mo and, further, the WSe$_{2}$ states are also tunable in the Mo-based heterostructures. Both the observations have their roots in the extended nature of the Mo-4$d$ states. 

Together, these observations suggest possibility of realizing an optical spectrum in these heterostructures whose tunability can be enhanced by exploiting the interplay between magnetic correlations and localized/extended TM orbitals. Molecules can host spatially localized excitons often referred to as Frenkel excitons~\cite{frenkelorig1,frenkelorig2,frenkelmolecule}.  They can also host ``charge-transfer'' excitons of Zhang-Rice~\cite{zhang1988effective} character. By contrast, the TMDs are known to host \textcolor{black}{delocalized Wannier-Mott~\cite{wannier} excitons}. In a recent work, we have shown that in a 2D-magnet/TMD heterostructure (CrI$_{3}$/WSe$_{2}$) the proximity magnetic field generated by the 2D magnet spin splits the Wannier excitons in the TMD. We also showed an emergent rich excitonic spectra in the heterostructure that contains the unperturbed Wannier and Frenkel excitons from the individual layers (TMD and the magnet respectively) and a new inter-layer exciton with charge-transfer character. One crucial observation in the  CrI$_{3}$/WSe$_{2}$ heterostructures was related to the energies and alignments of the excitons: the energies and relative positions of the intra- and inter-layer excitons were robust across all heterostructure variants and did not depend on number of layers and different (antiferromagnetic and ferromagnetic) magnetic interactions between the layers. While the novel inter-layer exciton with mixed Frenkel and Wannier character is interesting, \textcolor{black}{in consequence of a rigid local moment of these systems} very little tunability of the overall excitonic spectrum is available~\cite{mushir2025}. That is not the case with the molecule-TMD heterostructures we are studying here.  \textcolor{black}{Fig.~\ref{fig:frenkelvsct} compares a Frenkel exciton in CrI$_3$ to an exciton in the  molecule VO(dmit)$_{2}$.  The former mostly comprises an onsite $d^\mathrm{occ}{-}d^\mathrm{unocc}$ transition on the Cr, and is localized around it.  The latter is extended throughout the molecule, with essentially no onsite character.}

\begin{figure}[ht]
\includegraphics[width=18cm]{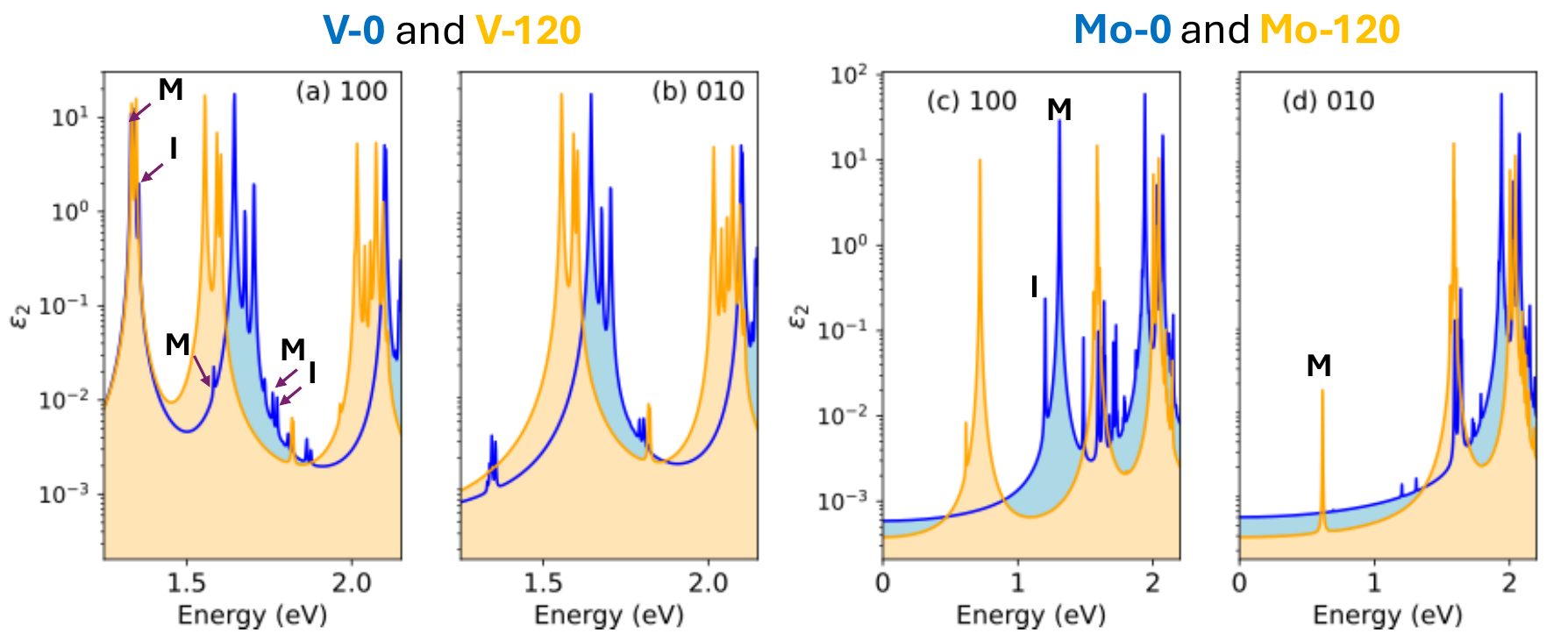}\\
\includegraphics[width=10cm]{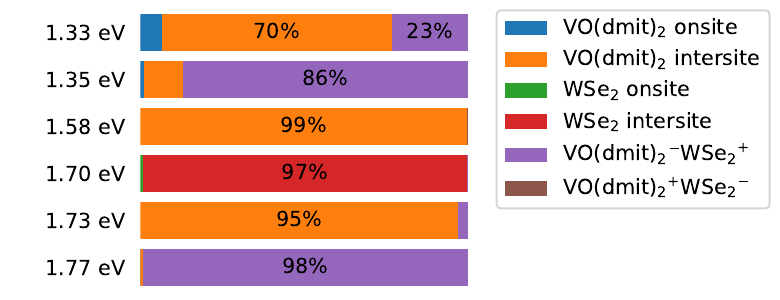}
    			\caption{The macroscopic dielectric response $\epsilon_2$ are plotted for different structural configurations of V (a,b) and Mo (c,d). \textcolor{black}{100 and 010 refer to the \textit{x} and \textit{y} components of the longitudinal elements of $\epsilon_2$.} Series of intra-layer and inter-layer excitonic peaks are observed. Molecular intra-layer, WSe$_{2}$ intra-layer and inter-layer excitons are noted respectively as M, W and I. A uniform optical broadening of only 1 meV is used at all energies and optical spectra are plotted in log scale to reveal all the weaker excitonic structures. In the horizontal bar plots the decomposition of the excitonic wavefunctions into different intra- and inter-layer components are shown. In the legend for the bar plots, `onsite' resolves the component of the exciton wavefunction that originates from purely atom-local onsite (Frenkel) transitions. It only makes an extremely weak appearance in the molecular 1.33 eV exciton. \textcolor{black}{The smallness of this component is strikingly different from the 2D magnet/WSe$_2$ systems, and indicates these excitons have essentially no Frenkel character. WSe$_{2}$ primarily hosts intersite transitions and can not have much of onsite transitions and this implies the green color is nearly absent from all bar plots.}  \textcolor{black}{The `intersite' component links electron-hole pairs located at different sites.  This includes optical transitions connecting the magnetic element and its ligands within the molecule (orange); or transitions between W and Se at different sites (red). VO(dmit)$_{2}$$^-$WSe$^+$ and VO(dmit)$_{2}$$^+$WSe$^-$ refer to excitons where the hole resides on the molecule and the electron in WSe$_2$, or vice versa. The bar plots show that to a good approximations excitons can be classed as purely WSe$_2$ (W), purely molecular (M), or purely interlayer (I).}}
	\label{fig:opt}
\end{figure}

As shown in Fig.~\ref{fig:opt} we observe several excitonic transitions in V-0 and V-120.  Horizontal bar plots (showing the wavefunction decomposition of the excitons in different atomic channels) in that figure establish that excitons can largely be divided into as purely WSe$_2$, purely molecular, and purely interlayer types. The peaks are largely grouped around two separate energies, near 1.3 eV and 1.7 eV. We observe two peaks at 1.33 eV and 1.35 eV with significant molecular components to their wavefunctions. The 1.33 eV peak can be called a molecular exciton while the 1.35 eV peak is an inter-layer exciton with the hole and the electron shared between the WSe$_{2}$ and molecular layer respectively. On analyzing the 1.33 eV molecular exciton, we find that the exciton does not have any dipole-forbidden onsite $dd$ or $pp$ components and hence is bright. Rather, the molecular exciton has a perfect analogy with the Zhang-Rice (ZR) excitonic picture~\cite{zhang1988effective} where the electron and hole are shared between the transition metal and ligand sites. Note that these excitons are often called charge-transfer excitons in the molecular exciton literature. This $pd$ or $sp$ charge transfer character of these excitons set them apart from to the ground state excitons in 2D magnets, like CrX$_{3}$~\cite{grzeszczyk2023strongly,wu2019physical,acharya2022real}, CrBrS~\cite{datta2025magnon,paulina,shao2025magnetically} and NiPS$_{3}$ \textcolor{black}{(see Fig.~\ref{fig:frenkelvsct} for a comparison of the two types).} In the latter cases the deeply bound excitonic states have a large onsite $dd$ ($\sim$70\%) character with secondary contributions ($\sim$15\% each) from intersite $dd$ and $dp$ electron-hole transitions, making them strongly Frenkel like. This is of remarkable significance since this implies that these molecular excitons are both brighter and more tunable than 2D magnetic counterparts excitons with similar binding energies. 

Before we explore the tunability of these molecular excitons, we study the overall exciton spectrum as generated by our theory. As expected, we observe the A and the B excitons from the WSe$_{2}$ layer at 1.70 and 2.07 eV respectively. This is primarily used as a convergence check for our essential results in terms of the density of k-mesh and the number of valence and conduction states (active space for excitonic correlations) that are included in the two-particle Hamiltonian that is solved to compute the excitonic eigenvalues and the eigenfunctions. In addition, we find intra-layer molecular excitons at 1.58 eV and 1.73 eV (see Fig.~\ref{fig:opt} (a)) and an inter-layer exciton at 1.77 eV. The small energy differences (the highest separation being 40 meV) between the molecular exciton and the inter-layer excitons may make their experimental detection difficult. There are further complications involved in resolving the brightness (intensities) of these peaks. The molecular exciton around 1.3 eV is bright and oriented along 100 (contrast Fig.~\ref{fig:opt} (a) with (b)). The 1.33 eV molecular exciton sits right next to the inter-layer exciton at 1.35 eV. Together, they almost appear as one bright peak which is a combined effect of the dipolar character (see Supplementary materials) of the molecular exciton and the charge transfer from the WSe$_{2}$ layer. The 1.58 eV molecular exciton, in contrast, is much darker. The WSe$_{2}$ A1s exciton is significantly bright and next to this bright peak are the much darker molecular exciton (+30 meV away) and the inter-layer exciton (+70 meV away). These two excitons are at least two orders of magnitude darker than the A1s exciton. Similarly dark inter-layer excitons are observed in TMD heterobilayers~\cite{yu2015anomalous,baranowski2017probing,rivera2018interlayer,wu2018theory} and and CrI$_{3}$/WSe$_{2}$ interfaces. We also observe the A2s exciton around 1.83 eV as was reported before~\cite{he2014tightly}, which is darker than the primary A1s exciton by roughly two orders of magnitude and is slightly darker than our observed interlayer exciton at 1.77 eV.       

To see how the V-0 excitonic spectrum can be tuned, we turn to V-120 (see orange curves in Fig.~\ref{fig:opt} (a,b)), where the V magnetic moment is 20\% smaller. We observe that the WSe$_{2}$ A1s exciton redshifts by $\sim$90 meV in V-120 compared to V-0. This is a direct consequence of the lower ionic vanadium character (smaller V moment) in V-120, extended wavefunctions and larger dipoles across the interlayer which are able to screen the WSe$_{2}$ electron-hole interaction moderately. However, the separation between the WSe$_{2}$ A and B excitons remains at $\sim$400 meV, same the value as that of the pristine WSe$_{2}$ layer. The dark molecular exciton state redshifts by $\sim$70 meV to 1.51 eV. The isotropic (in 100-010 plane) excitonic structure around 1.7 eV and the anisotropy around 1.3 eV remains a key distinguishing feature of the spectrum which is also helpful in separating the more molecular-like excitons from the WSe$_{2}$ excitons. 

We now shift our focus to Mo-0 and Mo-120 where the local magnetic moments are much smaller (at least by a factor of 2-3) compared to the V moments. In Mo-0, The WSe$_{2}$ A1s and B excitons appear at 1.63 and 2 eV respectively and molecular exciton appears at 1.3 eV. \textcolor{black}{The 1.63 eV exciton is the analog of the 1.7 WSe$_{2}$ exciton in VO(dmit)$_{2}$ heterostructure.} While these excitons bear similarities to V-120, the remarkable finding of our work is the clear energy separation of the inter-layer exciton from the rest of the manifold. We observe a strongly redshifted inter-layer exciton at 1.2 eV which is split of from the 1.3 eV molecular exciton by at least 100 meV (see Fig.~\ref{fig:opt} (c)), making it easier to detect and control in experiments. Both the 1.2 eV inter-layer and 1.3 eV intra-layer excitons have dipoles along (100).  As the Mo moment reduces further in Mo-120 and $\Delta_{dd}$ reduces, we observe that the molecular exciton strong redshifts into the telecom window and appears around at 0.75 eV (see Fig.~\ref{fig:opt} (d)). The fact that such strong redshift of the molecular exciton can be realized purely by replacing the active TM site and by making heterostructure geometries with varying angles between the layers, make this extremely attractive for devices design. Note that all of these functionalities were absent from both the TMD heterobilayers and the TMD/2d-magnet interface. In the TMD heterobilayers both the layers lack an explicit magnetic site and the only degree of freedom that could be tuned is  valley/pseudospin. On the other hand, in the TMD/2D-magnet heterostructure, while an active magnetic site is present  the electronic wavefunction is highly localized making transfer of charge and spin across the interface difficult.

In summary, we establish that excitons in molecule/TMD structures are highly tunable. Our theory shows that the band-edge features, their atomic-orbital characters and the $d^\uparrow$-$d^\downarrow$ splittings  can be controlled by varying angles between layers and by transition-metal-ion replacement. This has the potential for greatly enhancing their functionality, compared to the TMD heterobilayers or TMD/2-D magnet heterostructures. We pin down the source of this tunability to the distinct and yet delocalized character of the electron and hole excitations on both sides of the interface. A 2D magnet realizes a highly localized excitonic spectrum with mostly onsite $dd$ character, while excitons on a molecule containing a magnetic ion are extended with significant charge transfer character. The charge transfer nature of the molecular wavefunctions, in strong contrast to the atom-local character of the wavefunctions in 2D magnets, imply that their screening environment can be easily modified in presence of the TMD layers, a mechanism absent from the TMD/2D-magnet heterostructure. This inverse proximity effect is able to modify both the one-particle and excitonic features of the molecular excitons and the inter-layer excitons adding to the enhanced functionality of the molecule/TMD heterostructures. Further, we observe that by replacing the magnetic ion in the molecule from one with partially filled 3$d$ states to another with partially filled 4$d$ states adds to the functionality where the inter-layer exciton can be split off from the molecular excitons by $\sim$100 meV (which is easily resolvable in most optical measurements) and their brightness can be enhanced, making their detection easier. In the process we also observe the excitonic spectrum to smoothly redshift into the telecommunication window from the near visible-infrared window providing a possible realization of a quantum transduction in a molecular heterostructure.      
\section{Method}
\textit{Computational methodology:} The Quasiparticle Self-Consistent GW approximation~\cite{qsgw,questaal_paper} is a self-consistent form of Hedin's GW approximation. In contrast to conventional $GW$ implementations, QS$GW$ modifies the charge density and is determined by a variational principle~\cite{variational}.  A great majority of discrepancies with the experimental band gap can be traced to omission of electron-hole interactions in the RPA polarizability.  By adding ladders to the RPA, electron-hole effects are taken into account.  Generating \textit{W} with ladder diagrams has several consequences; most importantly, perhaps, screening is enhanced and \textit{W} reduced.  This in turn reduces fundamental band gaps and also valence bandwidths.  The importance of self-consistency in both QS$GW$ and $\mathrm{QS}G\hat{W}$ for different materials have been explored \cite{acharya2021importance}.

\section{Data Availability} 
The data that support the findings of this study are available from the corresponding author upon reasonable request.


\section{Acknowledgments}
This work was authored by the National Renewable Energy Laboratory for the U.S. Department of Energy (DOE) under Contract No. DE-AC36-08GO28308. Funding was provided by the Computational Chemical Sciences program within the Office of Basic Energy Sciences, U.S. Department of Energy. We acknowledge the use of the National Energy Research Scientific Computing Center, under Contract No. DE-AC02-05CH11231 using NERSC award BES-ERCAP0021783 and we also acknowledge that a portion of the research was performed using computational resources sponsored by the Department of Energy's Office of Energy Efficiency and Renewable Energy and located at the National Renewable Energy Laboratory and computational resources provided by the Oakridge leadership Computing Facility. The views expressed in the article do not necessarily represent the views of the DOE or the U.S. Government. The U.S. Government retains and the publisher, by accepting the article for publication, acknowledges that the U.S. Government retains a nonexclusive, paid-up, irrevocable, worldwide license to publish or reproduce the published form of this work, or allow others to do so, for U.S. Government purposes. %
The authors declare no competing interests.

\end{document}